\documentclass[letterpaper]{article} 
\usepackage{aaai25}  
\usepackage{times}  
\usepackage{helvet}  
\usepackage{courier}  
\usepackage[hyphens]{url}  
\usepackage{graphicx} 
\usepackage{natbib}  
\usepackage{caption} 
\usepackage{algorithm}
\usepackage{algorithmic}
\usepackage{xcolor}
\usepackage{subfigure}
\usepackage{hyperref}
\usepackage{newfloat}
\usepackage{listings}
\DeclareCaptionStyle{ruled}{labelfont=normalfont,labelsep=colon,strut=off} 
\floatstyle{ruled}
\newfloat{listing}{tb}{lst}{}
\floatname{listing}{Listing}
\usepackage[T1]{fontenc}

\title{\$100,000 or the Robot Gets it! Tech Workers' Resistance Guide:\\Tech Worker Actions, History, Risks, Impacts, and the Case for a Radical Flank}
\author{
Mohamed Abdalla}

\affiliations{
 University of Alberta \\
mabdall2@ualberta.ca
}

\usepackage{bibentry}

\begin{document}
\nocopyright
\maketitle

\begin{abstract}
Over the past decade, Big Tech has faced increasing levels of worker activism. While worker actions have resulted in positive outcomes (e.g., cancellation of Google's Project Dragonfly), such successes have become increasingly infrequent. This is, in part, because corporations have adjusted their strategies to dealing with increased worker activism (e.g., increased retaliation against workers, and contracts clauses that prevent cancellation due to worker pressure). This change in company strategy prompts urgent questions about updating worker strategies for influencing corporate behavior in an industry with vast societal impact. Current discourse on tech worker activism often lacks empirical grounding regarding its scope, history, and strategic calculus. Our work seeks to bridge this gap by firstly conducting a systematic analysis of worker actions at Google and Microsoft reported in U.S. newspapers to delineate their characteristics. We then situate these actions within the long history of labour movements and demonstrate that, despite perceptions of radicalism, contemporary tech activism is comparatively moderate. Finally, we engage directly with current and former tech activists to provide a novel catalogue of potential worker actions, evaluating their perceived risks, impacts, and effectiveness (concurrently publishing \textit{``Tech Workers' Guide to Resistance''}\footnote{``\textit{Tech Workers' Guide to Resistance}'' can be found at \url{https://www.cs.toronto.edu/~msa/TechWorkersResistanceGuide.pdf}
 or \url{https://doi.org/10.5281/zenodo.16779082} }). Our findings highlight considerable variation in strategic thinking among activists themselves. We conclude by arguing that the establishment of a radical flank could increase the effectiveness of current movements.
\end{abstract}

%

\begin{figure}[t]
\centering
   \includegraphics[width=\columnwidth]{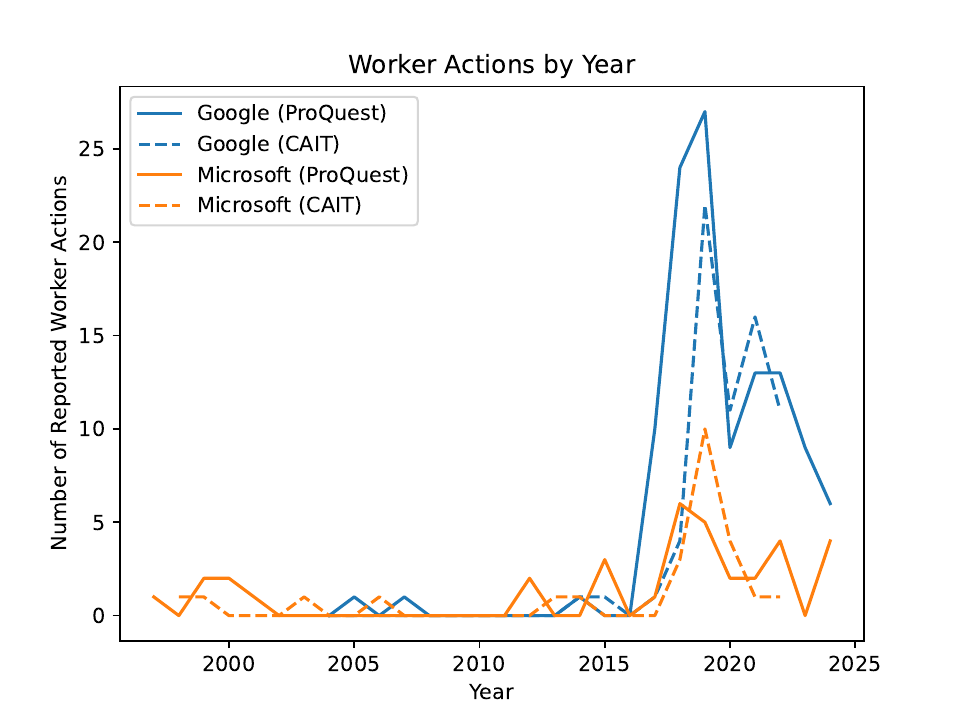}
\caption{Line plot of the number of actions by Google and Microsoft workers reported by major U.S. Daily Newspapers (as captured by ProQuest) per year from the late 1980s to March 22nd, 2025 and by Collective Action in Tech \cite{collectiveactionintech} from the late 1980s until 2023. }
\label{fig:numberactions}
\vspace*{-3mm}
\end{figure}

\section{Introduction}
Tech Workers at large U.S. technology corporations (i.e., Big Tech)\footnote{In this work, we use collapse \citeauthor{selling2023liberal}'s typology of workers: we consider all employees, contractors, and leaders as tech workers though many of our analyses focus on persons with high levels of technical talent. We limit our study to those who work in `Tech Firms', particularly Alphabet (henceforth: Google) and Microsoft.} are often portrayed as solitary, highly individualistic, and apolitical \cite{hill2024apolitical,lessons2019sixties,fred2018turner,fantasy2014land}. While low unionization rates \cite{rareunion2021} relative to other U.S. sectors and a historical reliance on individual scarcity over collective solidarity \cite{cory2025talk} have long defined the tech sector, including AI researchers \cite{ray2019demis}, recent headlines trumpet a shift, showcasing employees mobilizing for causes they champion \cite{usa2019talk,wakabayashi2018google,scheiber2020great}. 

The snowballing hype around AI, massive financial investments, and intensifying competition between major technology companies is projected to dramatically amplify AI's societal impact, thereby elevating the importance of worker activism. Unfortunately, compared to earlier instances of worker activism which were successful in quickly achieving their goals (e.g., cancellation of Google's Project Dragonfly and non-renewal of Project Maven) it appears that current efforts have been less effective. This is due, in part, to corporations adjusting their strategies to dealing with increased activism (e.g., retaliating against activists and putting in contract clauses that prevent termination based on worker pressure). This change prompts critical questions about the future of worker activism (both its nature and scope). What form of actions are workers taking and how should this change? How do the strategies employed by workers today compare to those employed by historical actions? What can workers today learn from those in the past?

Indeed, compared to the confrontational tactics employed by past labour and social movements\footnote{The title of this piece in homage to the action taken by NYU students who occupied a server room and held a ransom for the Black Panthers asking for “\$100,000 or the robot gets it!” \cite{julyk2008trouble,barron2015NYU}}, current tech activists appear hesitant to embrace action that risk being labelled violent, destructive, or overly disruptive.

This work aims to enhance the effectiveness of worker actions. We believe this can be achieved by: a) comparing how the frequency, forms, and evolution of contemporary worker actions compare against a historical inventory of activist tactics, and b) engaging with current and former tech activists to analyze the perceived risks, impacts, and effectiveness of various actions, thereby informing strategic decision making and education. 

Comparing historical and current approaches reveals that while tech worker actions are increasing in their number, their breadth remains narrow. We also observe that while there is substantial variation in perception of risk and impact of actions, there appears to be appetite for more radical actions. Combining these insights, this work concludes by arguing that the emergence of a ``radical flank'' could substantially increase the effectiveness of future tech industry activism.

\section{Defining Worker Action}
Past work exploring the activism of tech workers and AI researchers, has either focused on a handful of selected actions \cite{belfield2020activism}, or a specific type of action (e.g., collective action \cite{boag2022tech}). In contrast, our work takes a broader lens to what qualifies as an `action'. We define political action (henceforth shortened to action or act) as any purposeful action taken by a person or group of people to advance a political or social goal. Some examples of tech worker actions include:

\begin{itemize}
\item Jack Poulson, former senior Google scientist, quitting over ``forfeiture of our values'' regarding an effort to launch a censored search engine in China \cite{gallagher2018pouslon}.
\item Kevin Cernekee, former Google engineer, speaking to the press accusing the company of firing him for expressing conservative political beliefs \cite{ghaffary2019google}.
\item Frances Haugen, former Facebook data scientist, whistle-blowing to reveal that Facebook's own research showed that it amplifies hate, misinformation and political unrest. \cite{Pelley2021Facebook,olesen2025big}. 
\item Google employees staging sit-ins at company offices to protest cloud contracts with israel's government \cite{haskin2024sitin}.
\item Joe Lopez, former Microsoft engineer, interrupts CEO’s keynote with pro-Palestinian protest \cite{Bhuiyan2025lopez}.
\end{itemize}

As can be observed, there is variation in the number of actors (one vs many), of forms (from sit-ins to leaks), and of purpose -- no single political stance can tie them all together (though U.S. tech workers have been shown to lean towards the Democratic party in the U.S. \cite{vox2015politics}).

\section{Curating Historical Actions}
To curate a list of historical actions, we relied on two approaches. First, we relied on a study of relevant books \cite{malm2021blow,collinson2005resistance,rothstein2022recoding,harris2023palo}, articles/writings \cite{simplesabotagefieldmanualcia,sharp1973methods}, and academic publications \cite{boag2022tech,belfield2020activism}. All sources used for this section were read with the intention of translating actions from their historical context to the current tech setting. 

To complement the analysis of literature, we sought out expert input. Specifically, we engaged current and former employees of the Big Tech corporations (n=8) to support the identification and evaluation of potential worker actions. This study was approved by the University of Alberta REB (\#Pro00151408). In contrast to \citet{sharp1973methods}'s detailed list of individual actions, we opted for broad categories (e.g., classifying both disruptive workplace sit-ins and the occupation of a VP's office as `Disruptive Workplace Protests'). We chose this broader categorization for three primary reasons: (a) the creation of a truly comprehensive list of all specific worker actions presents a near-insurmountable challenge; (b) such a list would be too large to effectively analyze; and (c) we anticipate that a highly granular list would not retain its utility for future workers, a central aim of this study. Table \ref{tab:action_summary} presents a summary of the possible types of actions that can be taken by tech workers.

\subsubsection{Analyzing Historical Actions}
For each worker action, we relied on expert judgment to describe the action as positive or negative\footnote{Negative actions seek to halt or disrupt the normal operations of a company by refusal to obey orders or obstructing orders. Positive actions seek to build alternatives to existing systems or repairing relationships. This is discussed in more detail in the Discussion.}, and explicate the possible risks, impact, and overall effectiveness of the action. This process ensured that the analysis was grounded in both theoretical understanding and practical, expert-based knowledge. The analysis of each action can be found in the \textit{``Tech Workers' Guide to Resistance''}.

We grouped actions into higher level groups based on commonalities between them (e.g., all are some sort of protest, etc.). These groupings are not prescriptive, but rather a way to organize the data to make it easier both to search for actions and more quickly understand the variety available to workers.

\section{Studying Actions Taken by Tech Workers}
\begin{table*}[ht]
\centering
\begin{tabular}{|ll|} \hline

\multicolumn{1}{|l|}{\textbf{Quitting}} & \textbf{Lawfare}   \\ \hline
\multicolumn{1}{|l|}{Quit (quietly)}& (If fired) Sue \\
\multicolumn{1}{|l|}{Quit (internal protest)}   & Lawsuits   \\
\multicolumn{1}{|l|}{Quit (external protest)}   & Reporting to Regulatory Body\\
\multicolumn{1}{|l|}{}  & Engage with HR\\
\multicolumn{1}{|l|}{}  & Lobbying \\ \hline
\multicolumn{1}{|l|}{\textbf{Protesting}}   & \textbf{Malicious Compliance}  \\ \hline
\multicolumn{1}{|l|}{Public Protests}   & Minimal productivity (Work-to-rule)   \\
\multicolumn{1}{|l|}{Public Professional Protests}  & Collective delays  \\
\multicolumn{1}{|l|}{Internal Protests (active)}& Hiring lower quality candidates\\
\multicolumn{1}{|l|}{Internal Protests (passive)}   & Push for quitting  \\
\multicolumn{1}{|l|}{Disruptive Workplace Protests} & Elongate bureaucratic processes  \\
\multicolumn{1}{|l|}{Baiting Reactions} &  \\
\multicolumn{1}{|l|}{Striking} &  \\ \hline
\multicolumn{1}{|l|}{\textbf{Knowledge Transfer}}   & \textbf{Violence}  \\ \hline
\multicolumn{1}{|l|}{Leaking Anonymously}   & Towards Infrastructure \\
\multicolumn{1}{|l|}{Leaking Publicly}  & Towards Self   \\
\multicolumn{1}{|l|}{Publicize Actions} & Towards Others   \\
\multicolumn{1}{|l|}{Educational campaigns/teach-ins} &  Sabotage products \\
\multicolumn{1}{|l|}{} & Sabotage internal systems
  \\ \hline
\multicolumn{1}{|l|}{\textbf{Building Community}}   & \textbf{Positive Actions}  \\ \hline
\multicolumn{1}{|l|}{Organize}& Build Competing Products   \\
\multicolumn{1}{|l|}{Engage with activist/civil rights groups}& Create assistive tools/resources   \\
\multicolumn{1}{|l|}{Hiring Politically Aligned Staff} & Reperative Actions  \\ \hline
\end{tabular}
\caption{A summary of possible types of actions that can be taken by tech workers.}
\label{tab:action_summary}
\end{table*}

Despite the stereotype of tech workers being largely apolitical, there is historical precedence of tech/scientific workers often partaking in political actions. For example, it was a pair of Polaroid employees -- of which one was technical in nature (a chemist) -- who launched the first anti-apartheid boycott of a U.S. corporation \cite{dissent2020polaroid}. Likewise, in the late 1990s, Microsoft lost a class-action suit for incorrectly classifying employees as contractors \cite{microsoft1997WSJ}. Thus, our first course of action is to firmly establish the activity level of tech workers, and the types of actions undertaken. As mentioned previously, our exploration differs in 3 significant ways from prior work: a) we seek to be expansive in our coverage of the actions taken by tech workers, and b) we do not limit our analysis solely to collective actions, nor do we c) limit our analysis to non-violent actions.

\subsection{Data and Methods}
\begin{figure*}[t]
\centering 
\subfigure[]{\label{fig:a}\includegraphics[width=0.44\textwidth]{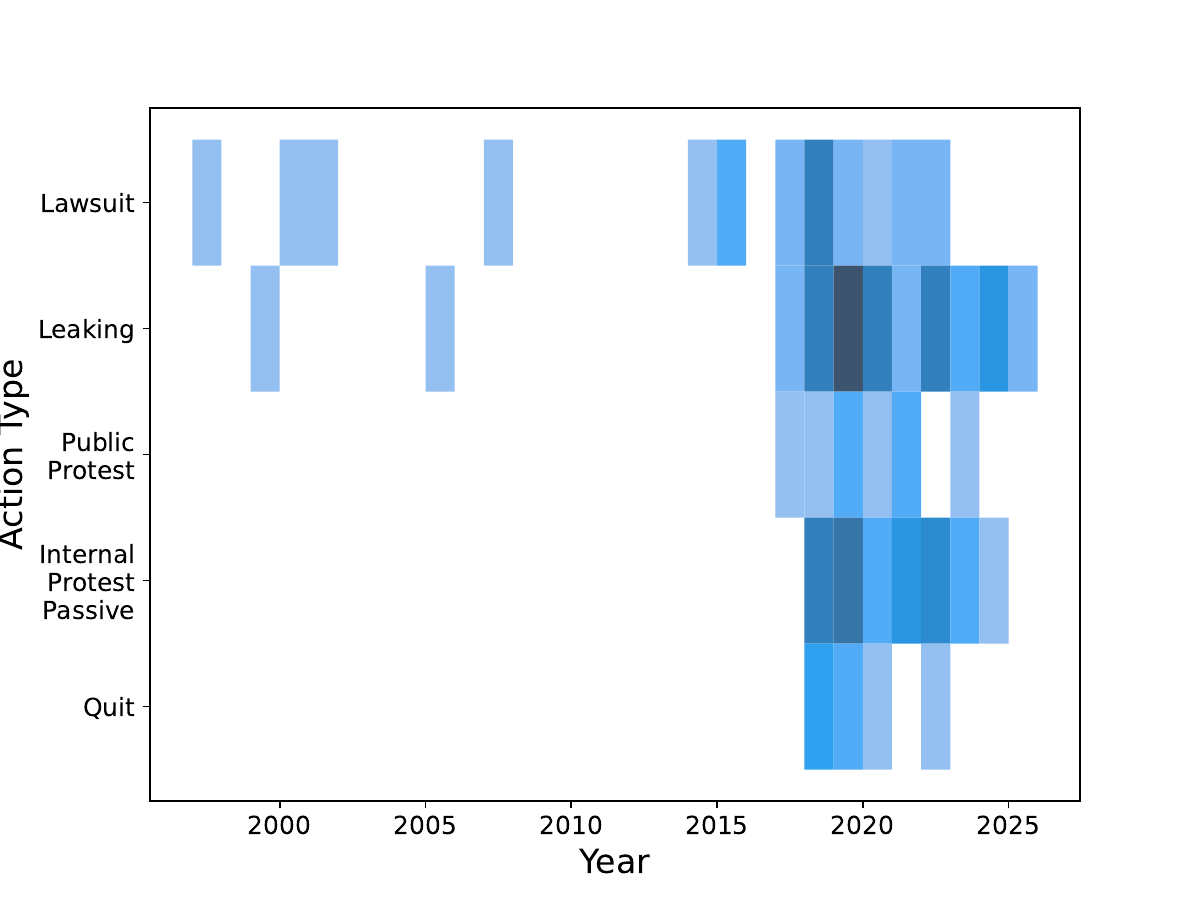}}
\subfigure[]{\label{fig:b}\includegraphics[width=0.44\textwidth]{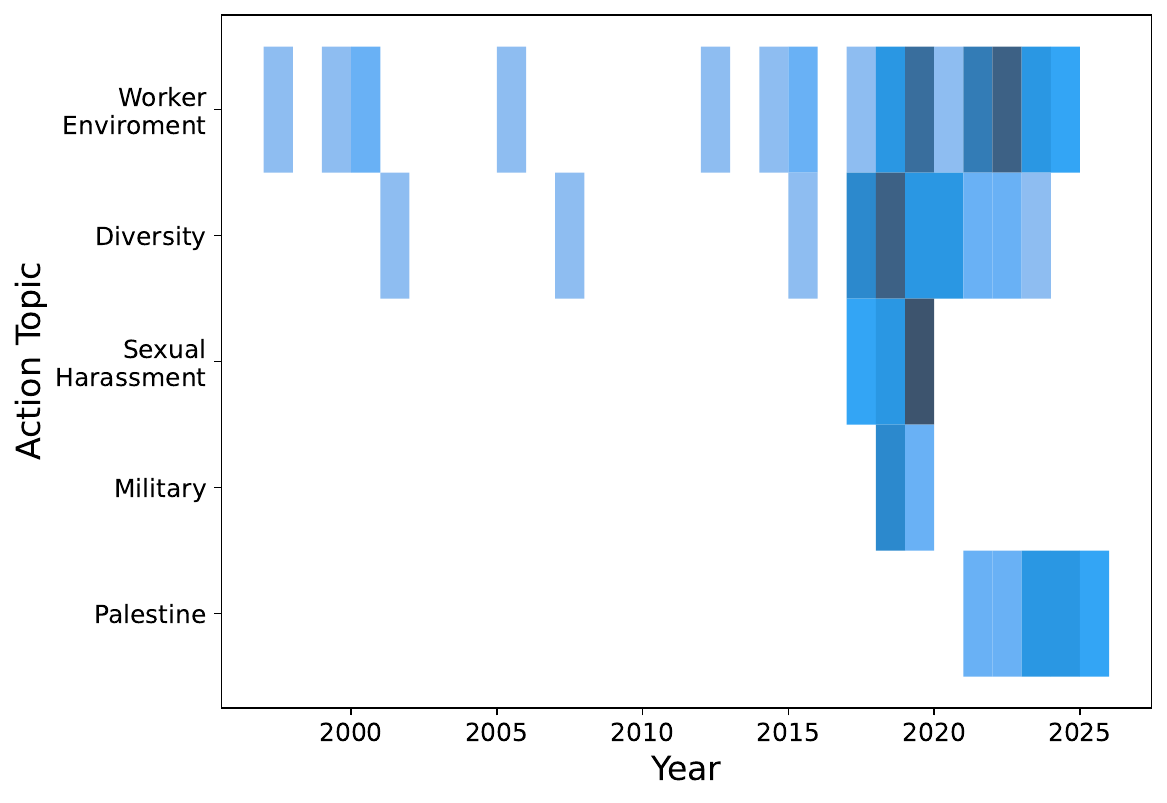}}
\caption{Histogram of the number of actions by (a) type and (b) focus/motivation as reported by major U.S. daily newspaper (as captured by ProQuest) per year from the late 1980s to March 22nd to 2025.}
\end{figure*}

To ensure a standardized and reproducible approach, we decided to leverage media coverage of actions at two firms (Google and Microsoft) for two reasons: First, these firms are amongst the most well-known and comprehensively-covered US tech companies in news media. Second, since we are manual filtering through articles, practical necessity compels us to limit the number of companies, accepting the reduced generalizability of such an analysis.

For this exploration, we relied on ProQuest's U.S. Newsstream \cite{proquest}. The U.S. newsstream is an archive of U.S. news content stretching back from the 1980s to present-day and contains full-text news sources of well-known national newspapers such as The New York Times and The Wall Street Journal, as well as over 80 regional and local news sources. We used the following query: 
\textit{(google OR alphabet OR microsoft) AND (worker OR employee OR contractor OR staff OR group)}. We limited our results' sources to full-text newspapers which were written in English and discussed the subject of `\textit{employees}' as defined by the ProQuest `Subject' filter. We conducted the initial search on March 22nd of 2025 and considered all results published before this date. 

Our query resulted in 11,859 news articles. We first used titles to identify relevant articles (i.e., those that discussed worker actions). For articles identified as possibly relevant (n=1,201), we read the full article to determine if they were truly relevant. For this limited subset  (n=154), we extracted all mentioned worker actions and associated dates. This process was conducted by two people and discussions were used to resolve disagreements.

We chose to capture our data from news coverage instead of using the  existing Collective Action in Tech (CAIT) dataset \cite{collectiveactionintech}, a database which documented collective actions from workers in the tech industry, because our scope is wider. CAIT focuses \textit{solely} on collective actions which is narrower than the scope of our paper: there are worker actions (e.g., quitting or whistle-blowing) that often are not taken in a(n explicitly) collective manner. An unintended benefit of this choice is that we are also able to explore the coverage of collective tech actions by the collection of major U.S. newspapers curated by ProQuest (as opposed to the search executed by \citet{nedzhvetskaya2021oxford} using NexisUni).

\subsection{Results}
\subsubsection{Actions over time}
Figure \ref{fig:numberactions} plots the number of actions taken by workers at Google and Microsoft over the past 40 years that have been reported by large U.S. daily newspapers. We can observe that the late 2010s represent the peak of reported worker actions. We can also see that Google employees tend to partake in more actions relative to Microsoft employees. 
Both Google and Microsoft workers were most active (as measured by coverage in news reports) 2015 and onwards, likely indicating a cultural shift of sorts or a change in the work of these companies. Uncovering the specific reason for this shift is important work that is beyond the scope of this paper. At both companies, we can see that after the flare-up pre-2020, there has been a relative down-turn in the number of worker actions. This could be caused by the increased crackdown on protests or for certain speech \cite{mohan2024gprotest,Sainato2024Google,mehta2025cisco}, or newspapers' choice to not report on these actions. This highlights an important limitation of this approach (editorial choice is a confounding factor which we are unable to correct for).

\subsubsection{Types of actions}
Tech workers employed a variety of strategies\footnote{Note: for this section we are using the framework developed in the previous section of the the paper.} to achieve their goals. Of the 153 actions observed in our dataset, 22\% (n=33) involved whistle-blowing/leaking to the media anonymously and an equal percentage, 22\% (n=33), were `Internal Passive Protests' (e.g., petitions, discussions on internal forums, etc.). This was followed by taking the employer to court at 16\% (n=25). We can see that the breadth of actions taken by tech workers has increased, Figure \ref{fig:a}. Before 2015, the most common actions were `Lawsuit' and `Leaking', whereas, after 2015, the most common actions were `Leaking' and `Internal Protest (Passive)' with the introduction of public protests and quitting. The full table can be found in Appendix Table \ref{tab:proquestactiontab_full}.

\subsubsection{Action Causes}
By far the most common cause is `Worker Environment' with nearly a third (28\%, n=43) of actions. We define `Worker Environment' to include concerns surrounding pay, benefits, and employment status. This was followed by 16\% (n=24) of actions relating to diversity or identity-based issues (e.g., race, gender, or age). Pro-Palestine and anti-war actions (if combined) take the third most common protest cause with 14\% (n=22) of all actions -- with pro-Palestine actions occurring at more than double the rate of general anti-war actions. `Sexual Harassment' has also been a large source (10\%, n=15) of actions. Like the types of actions, the motivation of actions has also changed over time. While desire for improvements in the work environment and diversity consistently motivate worker action, recent social changes and wider world events have precipitated worker actions, as seen in the actions that emerged following the peak of the \#metoo movement (2017-2018), the murder of George Floyd (2020), and israeli attacks on Palestine (2021, 2023--Present day (2025)). The full table can be found in Appendix Table \ref{tab:proquestactionmotiviation_full}.

\begin{figure*}[t]
\centering
   \includegraphics[width=0.7\textwidth]{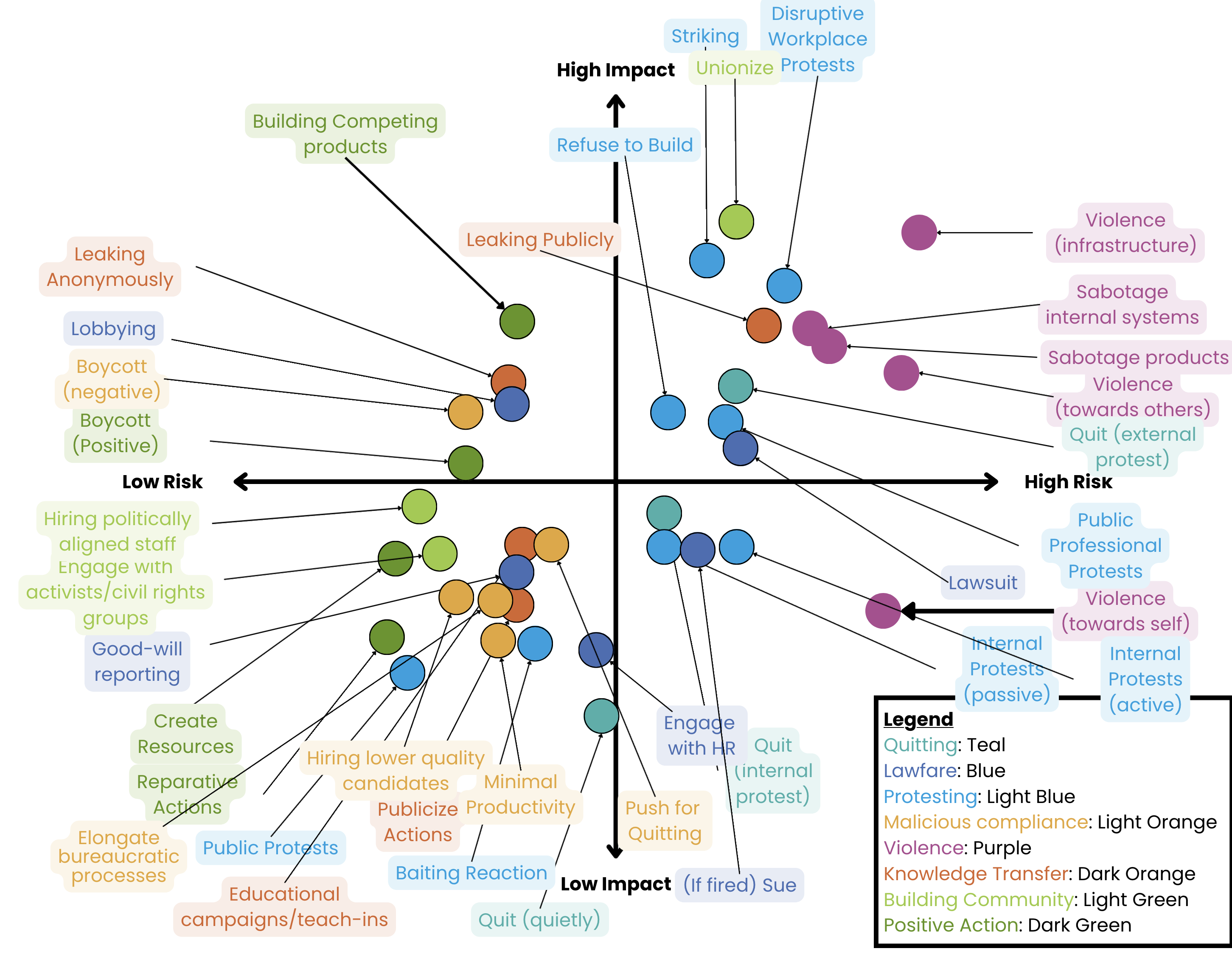}
\caption{Perceived general impact and risks of the full set of possible actions.}
\label{fig:riskimpactcartgrapgh}
\vspace*{-3mm}
\end{figure*}

\section{Quantifying Perceived Risk and Impact}
Having collated a broad set of actions, our aim was to ascertain expert perspectives on their potential impact and the accompanying risk. Is there wide agreement on impact or risk? Does high-impact necessarily entail high risk? Does the urgency of a cause affect the perceived impact of actions? For the sake of brevity in this section, we will focus solely on general impact; however, the same procedure was employed to quantify perceived risk and perceived immediate impact.

Instead of using a traditional Likert scale evaluation methodology asking experts to choose between a high/medium/low scale (which would require us to pre-define each category, thus possibly biasing the respondents), we instead used comparative annotations \cite{thurstone1927law,david1963method}. In its simplest form, comparative annotations can be built as paired comparisons which presents annotators with two options and asks them to choose which item is greater with respect to the property of interest (e.g., impact). With multiple questions, we would be able to generate an ordinal ranking of items. In this work, we use Best-Worst Scaling \cite{louviere1991best}, an advanced form of comparative annotations that expands the two-item comparative annotation task to multiple items. In this work, annotators are given n=4 items (i.e., actions) and asked to choose which action is most and least impactful. Each annotation of a 4-tuple provides us with five pairwise inequalities. For example, if among $a,b,c,$ and $d$, $a$ is selected as the most impactful and $d$ is selected as least impactful, then we know that $a>b, a>c, a>d, b>d,$ and $c>d$. From this, we can calculate an ordinal ranking of actions using a simple counting mechanism: the fraction of times an action was chosen as most impactful minus the fraction of times the item was chosen as least impactful \cite{Orme_2009}. Previous work has shown that reliable scores are obtainable from about $2N$ 4-tuples \cite{bws-naacl2016,kiritchenko2017best}.

\begin{figure*}[ht]
\centering
   \includegraphics[width=\textwidth]{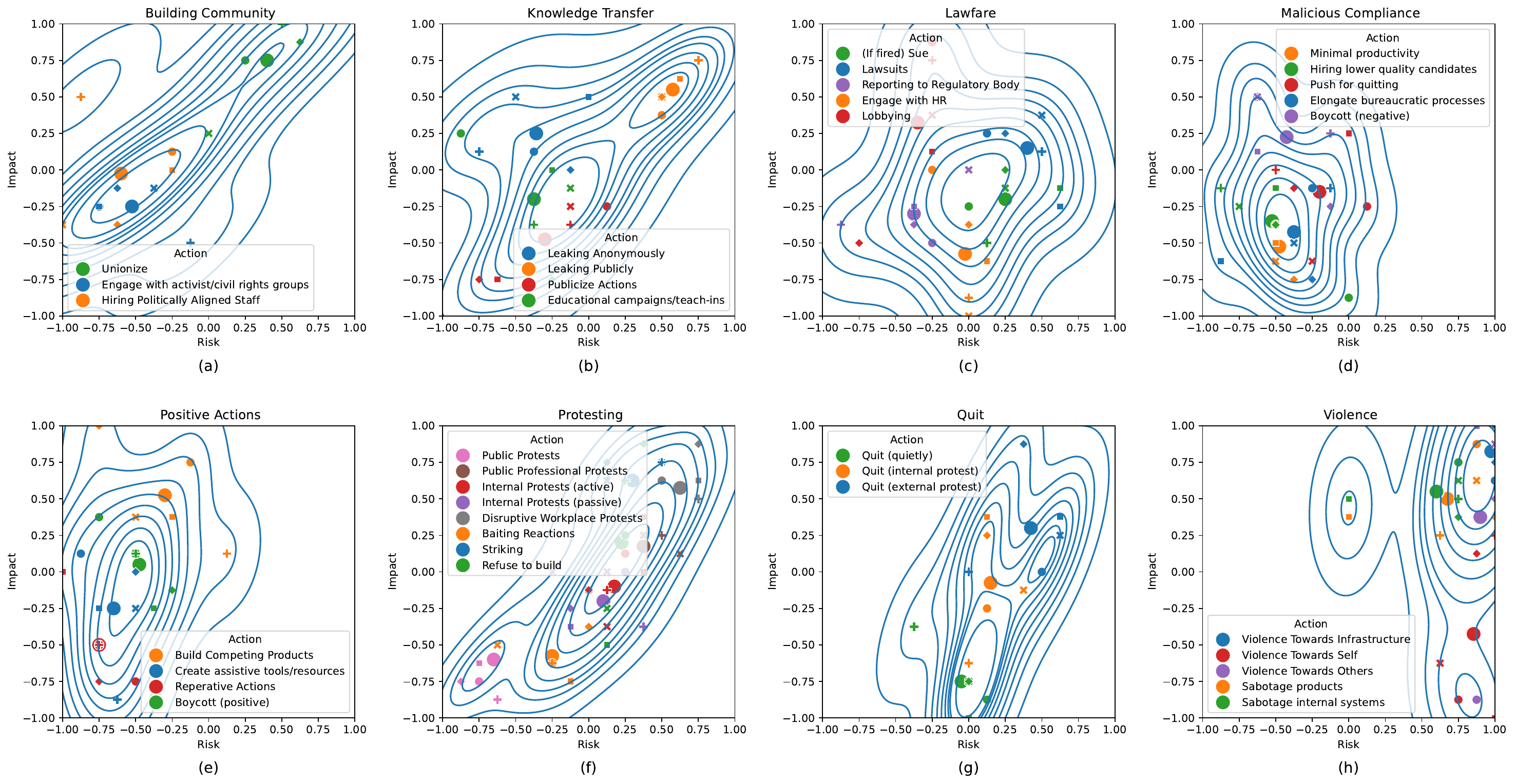}
\caption{Perceived general impact and risks of the full set of historical actions split by action category. Within each category, the large dots represent the overall annotation (i.e., considering all annotators) for each individual action. The smaller shapes represent an individuals rating for an individual action. Each annotator is given a consistent shape, and each action is given a consistent color. We used a KDE (Kernel Density Estimate) plot to visualize the probability density function of risk and impact (the blue lines). A larger version of this figure can be found in Appendix Figure \ref{fig:3x3actions}.}
\label{fig:3x3}
\vspace*{-3mm}
\end{figure*}

From our full list of 37 actions, we generated 74 unique 4-tuples\footnote{The tuples were generated using the BWS scripts provided by \citet{kiritchenko2017best}: \url{http://saifmohammad.com/WebPages/BestWorst.html}. This resource was also used to calculate the split-half reliability and the ordinal ranking of actions.} (each 4-tuple consisting of 4 distinct actions), with each action appearing in about 8 of the 4-tuples. Annotators were not provided detailed or technical definitions of impact or risk; rather, we encouraged annotators rely on their expertise and individual perspectives. All annotators for this task are all current or recently-former employees at Google and Microsoft who have participated in actions (either individually or as a collective). 5 annotators completed labeling the either set of 74 unique 4-tuples, with the remainder only annotating a small percentage.

\section{Results}
\subsection{Reliability of Annotations}
Since our comparative annotation procedure generates real-valued scores, we used \textit{split-half reliability} (SHR) to measure the quality of quality and reliability of our annotations \cite{cronbach1951coefficient,kuder1937theory}. SHR measures the degree to which repeated annotations of the same task would result in similar relative ranking of items. To measure SHR, for each 4-tuple, we split the 4-tuples into two bins. Each bin was used to independently produce the real-valued scores for each item which were compared using Spearman correlation. This process was repeated 1000 times. 

Our annotations had very high SHR: Spearman correlation (SpC) of 0.92 (i.e., `\textit{very strong}') for both rankings of perceived risk and impact. The high reliability of annotations, despite the lack of explicit instruction, indicates that there is a shared view in the perceived risk, impact, and immediate impact of tactics. Note: despite the generally high level of agreement on the ranking of actions, there were some strong individual differences for certain action (e.g., \textit{Violence Against Self}'s impact ranging from -1 to 0.25), Figure \ref{fig:3x3}.

\subsection{Risk vs Impact}
Figure \ref{fig:riskimpactcartgrapgh} plots, on a two dimensional grid, the perceived risk against the perceived impact of actions. Appendix Table \ref{tab:actionsquantified} presents the quantified scores for perceived risk, impact, and immediate impact for each action. There is generally a strong correlation between the perceived risk and the perceived impact of an action (SpC of 0.58) which increases further for risk vs immediate impact (SpC of 0.65). Figure \ref{fig:3x3} plots the perceived risk against the perceived impact of actions (large dots) split by action categories. For the five annotators who completed the full set of 74 annotations (370 binary comparisons), we place their personal ranking of actions resulting from their labels alone using smaller markers (e.g., small dots, x's, squares, etc.).

\subsubsection{Building Community} This family of actions exhibits a positive correlation between perceived risk and impact (SpC of 0.5). Generally, this family of action was viewed as having minimal risk (though unionization was considered high risk). Building community was rated as the second most impactful action generally, however, when seeking urgent impact, it dropped to the middle of the pack.

\subsubsection{Knowledge Transfer} This family of actions exhibits a moderate correlation between perceived risk and impact (SpC of 0.40). As a family of actions, it is among the most impactful even when urgent impact is needed, while having median risk. Within this family of actions, leaking (i.e., whistle-blowing) had high impact, and the impact (and risk) is increased further when done publicly. 

\subsubsection{Lawfare} Lawfare did not exhibit a strong correlation between perceived risk and impact (SpC of 0.20). As a family of actions, lawfare was amongst the lowest in terms of impact (dropping to lowest for immediate impact), while being near the median for risk to the worker. Within this family, lobbying was perceived to have the highest positive impact.

\subsubsection{Malicious Compliance} As a family of actions, the correlation between perceived impact and risk was weak (SpC of 0.30). Actions in this family are perceived to be the lowest impact (even for immediate impact) and near lowest risk.

\subsubsection{Positive Actions} Positive actions had a positive correlation between perceived risk and impact (SpC of 1.0). Actions within this family were deemed to have the lowest risk and medium impact (including for immediate impact). ``Building competing products'' stood out from the this family of actions as having high impact and relatively low risk.

\subsubsection{Protest} Protesting as a category of worker action has a very positive correlation between perceived risk and impact (SpC of 0.88). As a family, protesting was deemed to be somewhat impactful, rising to the second most impactful set of actions when immediate impact was needed. This family was perceived to be third riskiest. There is however, great variation (with respect to perceived risk and impact) within this family. Protests which are disruptive (e.g., `Striking' or `Disruptive Workplace Protest') were both the highest impact and the highest risk. The impact and risk of actions correlated with how tied to the message an action could be and how much it affected the workplace.

\subsubsection{Quit} Quitting had a strong correlation between perceived impact and risk (SpC of 1). Quitting was perceived to be the second riskiest family of actions, while being ranked in the lower half of impact (near median for immediate impact). There is a large range in impact despite a much narrower variation in risk. Quitting as a form of internal or external protest was seen to be much more effective than quitting quietly despite only a small change in risk.

\subsubsection{Violence} This family of actions exhibited only a weak correlation between perceived impact and perceived risk (SpC of 0.09). This category of actions was both the most impactful (for both long and immediate impact) and the highest risk. ``Violence against oneself'' was rated as the least effective action of this family.

\section{Discussions -- Morals, Strategy, Risks}
Thus far, we did not expound on what workers \textit{should} do. Comparing what has been done (\S 3) against what could done (\S 4) highlights the room for expanding the breadth of actions. The quantification of perceived risk vs impact (\S 5) can be used to determine where one's tolerance lies, and what actions a group can take without alienating its members. Unfortunately, strategic thinking is highly contextual; workers exist in different contexts, face different stressors and have different goals. For example, workers on visas face different risks than those who are citizens (regarding deportation for their activism as demonstrated by the recent deportations of pro-Palestine activists \cite{aljazeera_deport})\footnote{While this has traditionally held true, it seems that the governing  U.S. administration is considering deporting U.S. citizens \cite{rubin_2025_citizen}.}. Likewise, employees who are responsible for dependents are operating under different circumstances than those who don't have such a responsibility. 

Therefore, in this section, to enable workers to  assess for themselves how to prioritize actions, assess effectiveness, and develop coherent strategies, we discuss several commonly raised topics and questions related to worker activism. \textbf{It is crucial to note that this discussion and advice is purposefully kept at a high-level. Individuals must account for their specific contexts, risk tolerances, and legal jurisdictions.}

\subsection{Moral Injury and The Need to Act}
Moral injury refers to when ``there has been (a) a betrayal of `what's right'; and (b) either by a person in legitimate authority [...], or by one's self, (c) in a high stakes situation'' \cite{shay2014moral,litz2009moral}. Moral injury has been experienced by tech workers when they learn that the corporations for whom they work advance causes they are morally against (e.g., contributing to climate change \cite{stone_2024_oil}, or arming israel \cite{ibraheem2025kill,warren_2025_ibtihal}).

Self-reflecting on his own ethical decisions as a computer science graduate student, \citet{chan2020approaching} makes the distinction between moral obligations and moral aspirations. Moral obligations are actions whereby an individual becomes morally blameworthy should they fail to perform them (or in the case of a negative action: the blame is assigned when an action is performed). On the other hand, moral aspirations refer to actions where an individual is deemed morally virtuous for doing them, but not morally blameworthy for not performing them. He notes, for example, he has a moral obligation to not work on \textit{`building AI systems that enable genocide'}, but only a moral aspiration to pursue a research direction which maximizes \textit{`positive ethical impact'} \cite{chan2020approaching}. Through his examples, Chan highlights the wide variety of seemingly mundane choices made by graduate students that entail ethical consequences (e.g., selecting research directions, seeking funding, and mentoring).

Inspired by an Islamic paradigm, former Google software engineer Hasan Ibraheem arrives at a similar conclusion. Reflecting upon his work and his contributions to Google's `` direct support for the genocide against Palestinians due to their prioritization of profit over human lives'' \cite{ibraheem2025kill}, he concludes that despite not directly working on military technology, providing his labour to a company complicit in immoral actions makes him complicit. He suggests that workers should aim to quit or get fired in as loud a manner as possible as, per his piece, ``[o]rganizing does not absolve you of complicity indefinitely'' \cite{ibraheem2025kill}.

For those who feel strongly about particular issues such as Chan and Ibraheem, it's clear that contributing to such companies, even in tangential ways,  may still result in experiencing moral injury. When such injury cannot be avoided, moral injury can be counter-acted by taking actions which account to moral repair \cite{litz2009moral,walker2001moral}. Moral repair, a topic with rich literature, can be (overly) simplified to doing good deeds (more specifically appropriate amends) to regain positive self-judgment about one's self \cite{vives2023moral}. Continuing with Ibraheem's Islamic theme, this would be analogous to the Islamic concept of \textit{tawba} (which can be incompletely translated to repentance) \cite{abdullah2022role} where, in addition to remorse and abandonment of the action causing moral injury, the person should also seek to make amends for any harms they have caused \cite{yaqeen2022tawba}. We believe that any of the actions discussed in this work can help moral repair but this depends on a variety of factors (e.g., belief that there is net positive impact, which may be difficult to see).

Given the nature of this paper, we have prioritized the voices of workers who have been involved in worker actions or relevant discussions. This focus represents a sampling bias of sorts. It may be possible that most workers do not feel strongly about any particular issue and thus suffer no moral injury. In this case, to maximize impact of actions, it would be beneficial of workers to instill shared values into colleagues such that partaking in the system results in moral injury thus driving them to act (i.e., moral education).

\subsection{High-Level strategy}
While studying the differences between current and historical tactics is enlightening, it does not necessarily inform current and future tech activists on how to engage. Below, we start with a brief high-level vision of how actions should be approached.

First, activists should reflect on what drives them to act. Are they driven out of emotion? Guilt? Ego? While different people may have different source motivations ensuring clarity in one's motivation can lead to sustained activism \cite{jin2025stay,hall2019human}. What is the moral or political framework through which they see the world? Different frameworks will prioritize or disallow certain actions, and have a great effect in the balancing of priorities. What obligations and restrictions do you have placed upon you and how may they affect which actions you may partake in?

Then, it's important to consider the existing incentive structures for the systems you're trying to impact. The majority of humans and corporations are not (consciously) motivated by political ambitions or views. Rather, they react to incentives and dis-incentives placed upon them by the system in which they live and work. Understanding that system and the structures that drive decision making will enable activists to map out the power dynamics of their society and better simulate, situate, and execute for their cause \cite{bolsen_2025}. 

Lastly, activists should set their plans at the structural level. It's important to realize that all important causes require ``long term, tedious, frustrating, exasperating, boring, disappointing anti-climatic work over decades to achieve incremental gradual progress'' \cite{bolsen_2025,cait2025sabotage}.

\subsection{Violence vs Non-Violence}
It is now commonplace for those pushing for change to tout their adherence to non-violence as a mark of validity and superior moral standing \cite{dahlum2023moral}. For example, when describing Google workers' walkout to draw attention to Google's mishandling of sexual harassment claims and unequal pay, \citet{google2018walkout} highlights the civility and peaceful nature of the protest as a positive:
\begin{quote}
\textit{It helps that these faces are not angry, but mostly calm, and that the walkouts are largely peaceful and civilised. Things look very different to recent protests by Uber drivers, for example, who have staged more angry demonstrations, waving colourful signs and shouting slogans.} 
\end{quote}
Likewise, adherence to non-violence often extends to property destruction. As highlighted by \citet{malm2021blow}, past actions have gone so far as to requiring every participant to abide by a solemn pledge to not damage machines or infrastructure. Adherence to non-violent approaches also has academic proponents \cite{stephan2008civil,nepstad2011nonviolent} though much of this work has been attacked by later work for inaccuracies in analysis and cherry-picking of data \cite{Gerderloos2020deubnk}. This lionization of non-violence (and disregard for violent approaches) has also propogated within the tech activist community generally, and the AI ethics community specifically (e.g., through citations of \citet{stephan2008civil} by \citet{boag2022tech} as they focus solely on non-violent collective actions). While the view that non-violent actions are more effective than violent actions at achieving goals is mainstream\footnote{See section \textit{``The Need For a Radical Flank''} which discusses how, despite their significant contributions to the Civil Rights Movement, Martin Luther King Jr. is often lionized while el-Hajj Malik el-Shabazz (i.e., Malcolm X) is largely overlooked.}, this view is not uniformly shared:

\begin{quote}
\textit{It is strange and striking that climate change activists have not committed any acts of terrorism. After all, terrorism is for the individual by far the modern world’s most effective form of political action, and climate change is an issue about which people feel just as strongly as about, say, animal rights.} -- \citet{lanchester2007warmer}
\end{quote}

In fact, many activists and academics argue that violent acts (e.g., property destruction) can be, when engaged with appropriately, successful in directly achieving their goals or further enabling the success of non-violent actors \cite{malm2021blow} (see: ``The Need For a Radical Flank'' below). Historically, we have seen the suffragettes, the MK in South Africa, black activists in the U.S. and many other groups successfully use violence to work towards achieving their goals. More recently, we've seen the success of UK's Palestine Action in closing down more arms factories through direct violent action (sabotage) than over a year of popular mass protests \cite{palestineaction,palestineaction2}. 

As argued by \citet{malm2021blow}, most people are likely to agree that ``violence is \textit{prima facaie} bad'', but to disregard it as a matter of course in all circumstances is bad strategic thinking. A more realistic view on the use of violence is that it is a highly impactful tool that should only be used after intense strategic thinking that assesses the culture and contexts. For example, it was only after the perceived failure of non-violence that Nelson Mandela stated that ``we will have to reconsider our tactics. In my mind we are closing a chapter on this question of a nonviolent policy'' \cite{freedman2015mandela}. For a more recent example on the public support for violent action, consider the overwhelmingly bi-partisan support and sympathy for Luigi Mangione following his alleged assassination of a healthcare insurance company's CEO \cite{suciu2024social}.


\subsection{Public vs Private Actions}
Workers taking actions need to choose between advertising their action, doing it publicly without advertising, and doing it discreetly. Public actions increase scrutiny and possible personal repercussions, as well as the impact. Public actions can also serve to inspire others to act, the impact of which is likely much greater than the impact private acts that a single person could execute. Furthermore, the more people taking part in an action, the greater the impact and harder it will be to ignore or crush dissent. For example, it's trivial to fire and replace two protesting workers, but much harder to replace 2000. Likewise, fail-safes may prevent sabotage caused by an individual but would likely cease to function if there was wide participation from many workers.

At the same time, there can be many benefits to taking actions discreetly. If one is in a precarious position (e.g., on a visa or supporting multiple dependents), private actions allow them to positively impact their company while minimizing risk. There are also certain actions that would only work if executed discreetly. For example, actions which seek to surreptitiously drain the resources of companies (e.g., hiring lower quality candidates, hiring politically aligned staff, and pushing others to quit) would not be effective if exposed publicly.

\subsection{Positive vs Negative Acts}
Workers at companies tend to have different politics than the ownership class \cite{vox2015politics}. As such, their ability to influence the decisions made by management is often limited. Thus, workers often have to rely on disruptions and halting the normal operation of the company (i.e., negative actions). While negative actions have been very successful in the past, in a system driven to maintain the status-quo, such actions are not guaranteed to result in positive changes. That is, if all tech corporations partake in the same immoral activities, even if workers in all companies engage in actions, the likelihood of companies changing their activities is low.

Thus, to dramatically increase the cost of maintaining the status-quo, there should be companies that exist as moral alternatives to those being targeted by negative actions. Building these institutions is a very important positive action which we do not believe gets enough attention among tech workers. Instead of just quitting a Big Tech company to join another Big Tech company --- which can still be a courageous moral stance to take --- quitting a company to create a competitor with the intention of punishing them for their immoral action creates a real cost to their actions. Now instead of appealing to the morals of those who run companies and hide behind the excuse of share-holder value to maintain the status-quo, negative actions will incur a quantifiable cost that can be avoided.

\subsection{Changing from Within}
A common counter-argument used by those reluctant to quit is that, by being part of the system, workers will have more access to information, people, and knowledge that will make them more effective when pushing for change. That is, change can still be achieved from within a company.

For example, workers on the inside can benefit the movement through baiting and policing bad actors. Baiting colleagues who align with the harmful practices that the company practices and espouses into sharing things that directly violate company policy or local laws can result in these colleagues being reported, penalized, and forced to self-censor. The benefits of such internal pressure are two-fold: a) penalize bad faith actors while building community among those campaigning towards more moral company values, and b) eroding trust and harmony within the company culture. Internal actions can also cause coworkers to also feel morally obliged to act. Likewise, worker petitions were historically more impactful than petitions from outsiders.

On the other hand, many workers inside companies feel that, as small cogs inside a massive machine, their individual actions have little to no effect. Meaningful change is difficult as companies would not keep keep employees who pose any real threat to their business model or their actions.

Both views have truth in them. The loss of access (to information, people, and decision makers) can reduce the effectiveness of one's actions. Likewise, as companies notice the increasing politicization of their work-force, they continue to change internal policies, structures, and contracts \cite{perrigo_2024_nimbus} to effectively neuter any internal worker dissent. In such instances, the benefits of access are great diminished, and more drastic actions may be needed to effect change. 

\subsection{The Need For a Radical Flank}
Contemporary workers have demonstrably learned from the experiences of their predecessors. To sustain momentum, connections, and knowledge gained during periods of collective action (e.g., walkouts or protests), which historically tended to dissipate after an initial surge, workers have now established enduring organizations that facilitate knowledge transfer and continuity \cite{marx2024}. Consequently, the dismissal of individual active workers is less likely to halt movements entirely. Furthermore, workers have cultivated inter-company networks to foster solidarity and disseminate lessons learned \cite{marx2024}.

However, tech worker activism currently lacks a radical flank. While there exist radicals in tech (e.g., neo-luddites \cite{dries2024future}), they are not a flank of current activists. \citet{haines1984black} highlights how many past movements ``divide into `moderate' and `radical' factions during their development, having been observed in the U.S. labour movement, women's movement, anti-nuclear movement, and the black revolt in the United States''. In the literature, the radical flank adopts a more maximalist goal and more radical tactics \cite{tompkins2015quantitative}, thus presenting the moderate flank as a reasonable alternative \cite{malm2021blow}. 

Per social movement theory, the lack of factionalism within existing tech activist organizations (i.e., no radical flank), suggests this social movement as being in its infancy. This understanding can help workers situate themselves with the context of past struggles and enable them to assess which strategies are more appropriate. 

Second, understanding the inherent historical tension between moderate and radical flanks will enable them to fight united for a cause. The role of the moderates is not become radical. Instead, the moderate flank should pro-actively work to position itself to benefit from the spikes in interest resulting from radical actions \cite{fortier2025unraveling} while making sure not to preemptively undercut any of the actions of a radical flank by undue disparagement radical tactics. As described by \citet{malm2021blow} of the Black Civil Rights struggle, a jailed Martin Luther King ``could signal that if channels of non-violence remained closed, [...] negreos unquestionably will look to untried and perhaps less responsible leaders – notably Malcolm X'' (who represented the radical flank). He concludes that the civil rights movement's moderate flank was able to win the Act of 1964 because, relative to the radical flank, it appeared to be the lesser of two evils.

The literature on the effectiveness of radical flanks is mixed. \citet{tompkins2015quantitative}, studying the effects of violent radical flanks, finds that they resulted in increased repression but were not ``necessarily detrimental'' to progress. \citet{simpson2022radical}, like \citet{haines1984black}, found that radical flanks increase support and perception of moderate flanks. An issue with much of the academic literature on radical flanks is their conservative nature, framing the radical flank as doomed for failure and only existing to further the moderates. Furthermore, much of the literature focuses on violent radical flanks but radical flanks are not \textit{necessarily} violent.

The strategic question for the radical flank is how radical to be. Too radical and they risk discrediting the movement as a whole, a serious `negative radical flank effect' \cite{haines1984black}. Not radical enough, and they are ineffective moderates. The tech workers who compose the radical flank must balance pushing the movement forward and upping the stakes while staying close and relatable enough to the moderates -- a challenging balance to strike.

\section{Limitations}
This work has several limitations. Firstly, our investigation of tech worker actions focused on two corporations: Google and Microsoft which constitute only a fraction of the broader tech workforce, both domestically and internationally. Consequently, the results (i.e., numbers, motivations, and the forms of actions) may not be representative of the wider tech worker class. For instance, climate-related protests appear to be more frequently reported among Amazon employees than those at Google or Microsoft (though this may be also be due to editorial choices in coverage -- another limitation). This selective focus was a pragmatic decision necessitated by the considerable volume of data; our initial search for these two companies alone yielded nearly 13,000 articles.

Further, despite the large number of articles, our news archive (ProQuest U.S. Newsstream) does not offer exhaustive coverage of all news sites within the U.S. Consequently, actions covered exclusively in smaller publications are not captured by our search (e.g., a comparison with the dataset compiled by \citet{collectiveactionintech} revealed an overlap of approximately 40\% of the actions). Regrettably, access to a more comprehensive archive of news sources was not available to the authors. Nevertheless, we posit that the trends observed in our analysis likely hold true to some extent, given the significant alignment in patterns (as illustrated in Figure \ref{fig:numberactions}) between our findings and those from \citet{collectiveactionintech} who employed slightly different inclusion and exclusion criteria.

Furthermore, not all actions are publicized or covered by publications. Despite these limitations, we believe our current approach offers a sound methodology, consistent with prior research \cite{zippi2024united,zippi2023us,berdahl2023strategies}, and possesses methodological rigor, ease of replication.

Similarly, while we compile a large set of possible worker actions, we almost certainly missed actions; the literature on worker actions throughout all of history is massive and not easily digested. Unlike other enumeration of possible actions (e.g., \citet{sharp1973methods} who listed 198 actions), we purposefully propose abstract descriptions for action groupings (e.g., public protests can be marches, motorcades, parades, or religious processions -- all different actions per \citet{sharp1973methods}'s work). These broader action groupings enable us to draw conclusions that will be inclusive of all current and future variations in action types within a given category, preventing our work from being artificially limited or dated with irrelevant actions.

The purpose of quantifying perceived risk and impact was to understand current activists' perceptions of their strategic choices, not to serve as a universal guide for future strategy. Other activists could arrive at different risk vs impact tradeoffs. Furthermore, the perception of risk and impact is heavily affected by societal programming which may be artificially limiting. Nevertheless, we believe that this is a good mapping exercise which can be taken by activist groups to help expand their view on what actions are possible and palatable to their constituents.

In the same vein, for each action presented in the ``\textit{Tech Workers' Guide to Resistance}'', we presented our current rationale to facilitate strategic thinking about the goals and necessary planning. It is our hope that readers will use this resource to develop their own strategies and apply them effectively within their specific contexts.

Lastly, the academic literature which we rely on to argue for the creation of radical flank has many limitations. Most severe, nearly all of this literature frames the radical flanks role as solely to increase the probability of success for the moderates, \textit{a priori} discounting the possibility of success for the radicals. This pessimistic framing may be needlessly hampering the imagination and success of such movements. Second, many of the studies about radical flanks are done in ``democratic'' environments, which is fundamentally different from the workplace which is not democratic. This distinction will affect the strategic calculus. For example, radical flank action is sometimes criticized as undemocratic \cite{hayes2024malm}. However this strain of critique derives legitimacy from the claim that the overall system is democratic -- a claim which does not apply in corporations.  

\section{Conclusion}
Technology workers possess a diverse repertoire of actions to select from when advocating for change within corporations. As corporations change their strategies to deal with increased activism, activists should likewise consider a change in strategy. To enable this, this study analyzed the breadth and extent of worker actions at two large firms in comparison to historical precedents. While the variety of actions employed by workers to pursue change has expanded, it remains considerably narrower than the theoretical possibilities. We argue that expanding the scope and types of actions can increase effectiveness of campaigns.

To aid this process, we created ``Tech Workers' Guide to Resistance''. Employing a synthesis of literature review and expert insights, we developed an expansive list of resistance actions available to workers. For each action, we quantified the perceived risk and impact, and provided accompanying notes and illustrative examples (drawing from past tech worker actions where feasible).

\newpage
\section{Acknowledgments}
We are especially grateful for the workers who provided their valuable time and input to analyses presented in this work. This work would not have been possible without them. Mohamed is funded by the University of Alberta and a Canada CIFAR AI Chair. The views presented in this work is not necessarily those of the funders who did not have any direct role in the conceptualization or execution of this study.

\bibliography{aaai25}

\newpage
\appendix
\onecolumn
\section{Appendix}
\setcounter{figure}{0}
\setcounter{table}{0}
\begin{table}[htbp]
\centering
\begin{tabular}{lc}
\multicolumn{1}{c}{\textbf{Action}} & \multicolumn{1}{c}{\textbf{Number}} \\ \hline
Leaking Anonymously & 33  \\
Internal Protest Passive& 33  \\
Lawsuit & 25  \\
Leaking Publicly& 14  \\
Public Protest  & 12  \\
Unionize& 10  \\
Reporting to Regulatory Body& 9   \\
Internal Protest Active & 8   \\
Disruptive Workplace Protest& 8   \\
Quit (external protest) & 7   \\
Engage with HR  & 4   \\
Quit (internal protest) & 2   \\
Public Professional Protest & 2   \\
Violence Towards Self   & 1   \\
Reparative Actions  & 1   \\
Publicize Actions   & 1   \\
Building Assistive Tool & 1   \\
Engage with activist/civil rights group & 1   \\
Lobbying& 1   \\
Leaking Publicly & 1   \\
Strike  & 1   \\   \hline
\end{tabular}
\caption{Number of action types used by workers at Google and Microsoft in our dataset. \label{tab:proquestactiontab_full}}
\end{table}

\begin{table}[htbp]
\centering
\begin{tabular}{lc}
\multicolumn{1}{c}{\textbf{Topic}} & \textbf{Count} \\ \hline
Worker Enviroment  & 43 \\
Sexual Harassment  & 16 \\
Palestine  & 15 \\
Gender & 10 \\
Other  & 9  \\
Race   & 7  \\
Military   & 7  \\
AI Ethics  & 7  \\
Diversity  & 6  \\
Trump  & 5  \\
LGBT   & 5  \\
ICE& 5  \\
Censorship & 5  \\
Vaccines   & 3  \\
Faith  & 2  \\
Climate& 2  \\
Abortion   & 2  \\
Age& 1  \\
Corruptions& 1  \\
Policing   & 1  \\
Anti-trust & 1  \\ \hline
\end{tabular}
\caption{Number of action topics championed by workers at Google and Microsoft in our dataset. \label{tab:proquestactionmotiviation_full}}
\end{table}

\begin{table}[ht]
\centering
\begin{tabular}{l|ccc}
\multicolumn{1}{c|}{\textbf{Action}} & \textbf{Risk} & \textbf{Impact} & \textbf{Impact Immediate} \\ \hline
Violence Towards Infrastructure  & 0.975 & 0.825   & 0.925 \\
Unionize & 0.4   & 0.75& 0.025 \\
Striking & 0.3   & 0.625   & 0.75  \\
Disruptive Workplace Protests& 0.625 & 0.575   & 0.75  \\
Leaking Publicly & 0.575 & 0.55& 0.45  \\
Sabotage internal systems& 0.6   & 0.55& 0.75  \\
Build Competing Products & -0.3  & 0.525   & -0.075\\
Sabotage products& 0.675 & 0.5 & 0.575 \\
Violence Towards Others  & 0.9   & 0.375   & 0.4   \\
Lobbying & -0.35 & 0.325   & -0.125\\
Quit (external protest)  & 0.425 & 0.3 & 0.35  \\
Leaking Anonymously  & -0.359& 0.25& 0.4   \\
Boycott (negative)   & -0.425& 0.225   & 0 \\
Refuse to build  & 0.225 & 0.2 & 0.4   \\
Public Professional Protests & 0.375 & 0.175   & 0.375 \\
Lawsuits & 0.4   & 0.15& -0.25 \\
Boycott (positive)   & -0.475& 0.05& -0.125\\
Hiring Politically Aligned Staff & -0.6  & -0.025  & -0.2  \\
Quit (internal protest)  & 0.15  & -0.075  & 0.1   \\
Internal Protests (active)   & 0.175 & -0.1& 0 \\
Push for quitting& -0.2  & -0.15   & -0.075\\
(If fired) Sue   & 0.25  & -0.2& -0.375\\
Educational campaigns/teach-ins  & -0.375& -0.2& -0.2  \\
Internal Protests (passive)  & 0.1   & -0.2& -0.075\\
Create assistive tools/resources & -0.65 & -0.25   & -0.175\\
Engage with activist/civil rights groups & -0.525& -0.25   & -0.075\\
Reporting to Regulatory Body & -0.375& -0.3& -0.375\\
Hiring lower quality candidates  & -0.525& -0.35   & -0.5  \\
Elongate bureaucratic processes  & -0.375& -0.425  & -0.35 \\
Violence Towards Self& 0.854 & -0.425  & -0.1  \\
Publicize Actions& -0.3  & -0.475  & -0.3  \\
Reperative Actions   & -0.75 & -0.5& -0.5  \\
Minimal productivity & -0.475& -0.525  & -0.475\\
Baiting Reactions& -0.25 & -0.575  & -0.35 \\
Engage with HR   & -0.025& -0.575  & -0.475\\
Public Protests  & -0.65 & -0.6& -0.35 \\
Quit (quietly)   & -0.05 & -0.75   & -0.725   
\end{tabular}
\caption{Risk, Impact, and Immediate Impact as rated by our annotators. \label{tab:actionsquantified}}
\end{table}

\begin{figure}[htbp]
\centering
   \includegraphics[width=\textwidth]{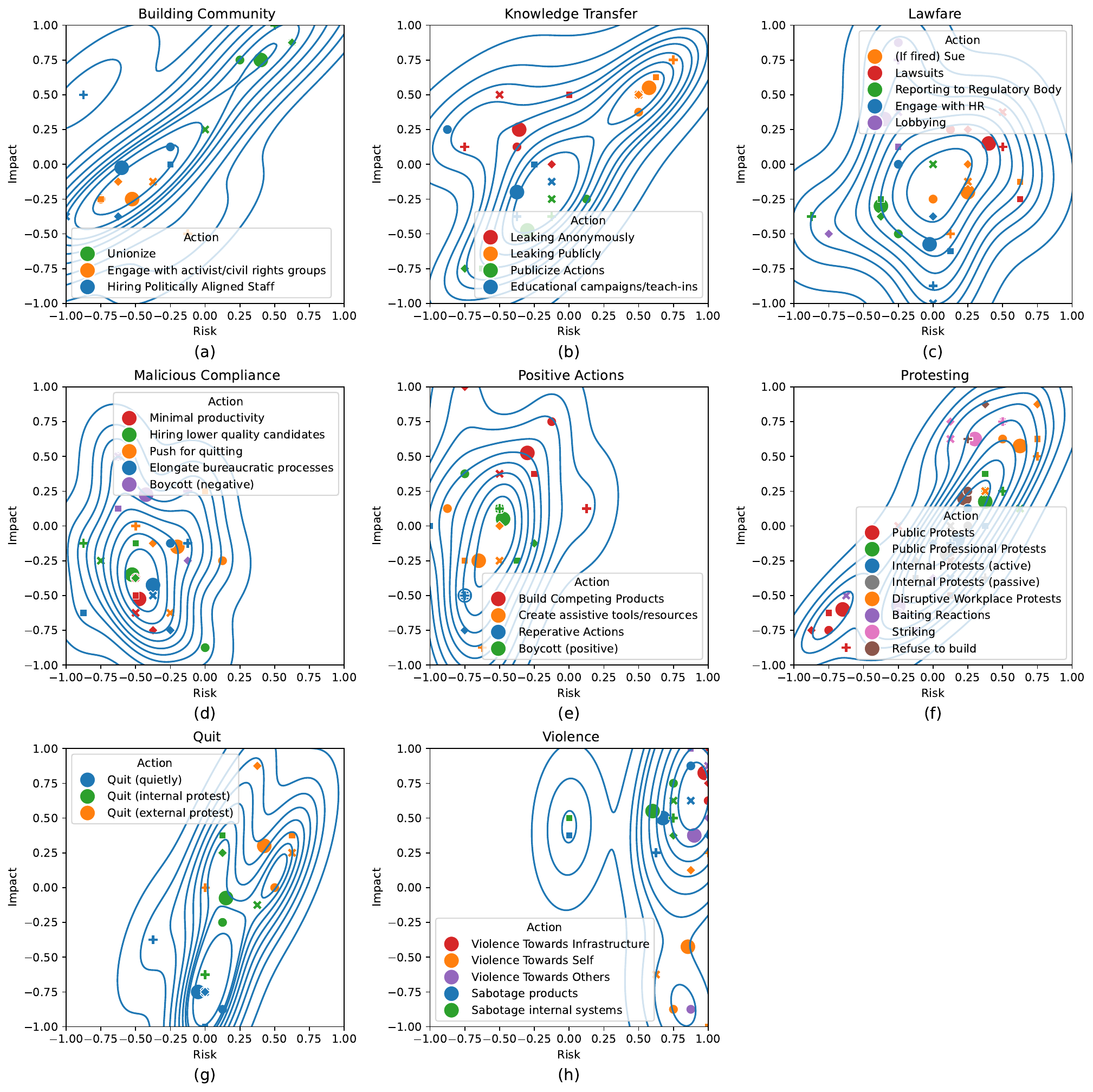}
\caption{Quantified relative perceived impact and risks of the full set of historical actions.}
\label{fig:3x3actions}
\end{figure}

\end{document}